\documentclass[%
 reprint,
 amsmath,amssymb,
 aps,
]{revtex4-2}

\usepackage{graphicx}
\usepackage{dcolumn}
\usepackage{bm}

\usepackage{amsmath,amssymb}
\usepackage{xcolor}
\usepackage{tikz}
\usetikzlibrary{
    arrows.meta,
    calc,
    decorations.pathreplacing,
    positioning
}
\begin{document}

\preprint{APS/123-QED}

\title{Hybrid Heralded Noiseless Amplification with Finite-Cutoff Quantum Scissors}

\author{Haixi Miao}
 \author{Timothy C. Ralph}
\affiliation{%
{Centre for Quantum Computation and Communication Technology, School of Mathematics
and Physics, University of Queensland, Brisbane, Queensland 4072, Australia}
}%

\date{\today}

\begin{abstract}
A noiseless linear amplifier (NLA) can probabilistically amplify an optical state
without the noise required by deterministic phase-insensitive
amplification. We study a hybrid
amplifier in which a finite-cutoff quantum-scissor NLA is placed
between two single-mode squeezers. Analysis based on an ideal (infinite cutoff) NLA finds a gain enhancement from the squeezing. We explore the physics of this enhancement as the cutoff of the quantum scissors is increased. The low-cutoff sequence shows how
this gain enhancement emerges from the
truncated Fock space. Cutoff 3 is the lowest order at which the
additional even- and odd-photon components can both contribute to
gain enhancement, albeit with some skewing of the coefficients which reduces the fidelity. 
As the cutoff is increased the fidelity improves. 
However, unlike
the ideal transformation, the finite-cutoff device depends
on the phase of the coherent amplitude relative to the squeezing
axes, with states aligned with the anti-squeezing requiring higher cut-offs to achieve high fidelity. We track behavior to high cutoffs and eventually see the performance predicted in the ideal theory emerge. 
\end{abstract}

\maketitle


\section{\label{sec:level1}Introduction }

Loss limits the distribution of optical quantum states and entanglement over long distances, and remains a problem in quantum communication \cite{briegel1998,weedbrook2012,dias2017}. Amplification may appear to provide a direct way to compensate for this loss. However, a deterministic phase-insensitive amplifier must add noise \cite{caves1982}. This noise preserves the field commutation relations and prevents the deterministic cloning of an unknown quantum state \cite{wootters1982}.

A noiseless linear amplifier (NLA) avoids this added noise by giving up deterministic operation \cite{ralph2009,walk2013,combes2016}. When the operation succeeds, it amplifies the displacement of a coherent state without increasing its noise. A measurement heralds each successful event, while the other outcomes are discarded. The ideal NLA is not physical over the full Fock space because its amplification factor grows without bound with photon number. An exact implementation would therefore have a vanishing success probability, and a physical device must restrict the transformation to a finite photon-number range.

Ralph and Lund proposed a linear-optical NLA based on quantum scissors, single-photon ancillas, and conditional photon detection \cite{ralph2009}. Quantum scissors were originally introduced to truncate optical states in the Fock basis \cite{pegg1998}. Several forms of probabilistic noiseless amplification have since been demonstrated experimentally \cite{xiang2010,ferreyrol2010,zavatta2011,kocsis2013,chrzanowski2014}. 
These results show that finite-dimensional NLA operations can be realised with current optical components. More recently Guanzon et al \cite{guanzon2022} have shown how to increase the cut-off of the quantum scissor NLA beyond the 1 photon cut-off of the original proposal. Increasing the cutoff improves the approximation to the ideal transformation, but requires more resources and generally lowers the success probability \cite{mcmahon2014,winnel2020,guanzon2022}.

The NLA has been studied for entanglement distillation, probabilistic correction of loss, quantum communication, and quantum-state cloning~\cite{ralph2011,haw2016}. Quadrature squeezing has also been proposed as a means of improving linear-optical NLA schemes~\cite{yang2013}. These studies establish the potential benefit of combining Gaussian operations with probabilistic amplification. Here, we focus on how this benefit emerges when the NLA is restricted to a finite Fock-space cutoff.

An ideal hybrid device consisting of a single-mode squeezer, an NLA, and a second squeezer was studied in Ref.\cite{mommers2017}. The squeezing operations allow the device to produce an effective gain larger than the gain of the NLA alone. A target gain can therefore be reached using a weaker NLA, suggesting a possible increase in heralding probability. However, this result assumes an ideal, infinite-cutoff NLA, which has zero success probability over the full Fock space. It does not show whether the same mechanism survives when the NLA is replaced by a physical finite-cutoff operation.

In this work, we examine how squeezing-assisted amplification emerges
as the NLA cutoff is increased. We show that cutoff 3 is the first
order at which the additional even- and odd-photon components can
produce gain enhancement in the squeezing direction selected by the
ideal transformation. We then test the higher Fock coefficients up
to cutoff 11 and identify an operating point at which their matching
conditions cluster around a common second-squeezing parameter. For a
phase-aligned coherent input, the resulting device reaches the target
gain using a lower internal NLA gain and therefore gives a higher
heralding probability than the corresponding bare NLA. However, we also show
that the finite cutoff introduces a dependence on the phase of the
coherent amplitude relative to the squeezing axes which remains significant until higher cutoffs. Finally, we apply
the device to one mode of a two-mode squeezed vacuum state and
identify the range in which the probability advantage is retained. However, at modest cutoffs the phase dependence of the device leads to reduction in fidelity relative to the ideal transformation.

\section{Hybrid squeezing amplification}
\label{sec:hybrid}
\subsection{Ideal noiseless linear amplification}

We first review the ideal hybrid squeezing amplifier following the approach of Mommers \cite{mommers2017}. The device consists of a single-mode squeezer, an ideal NLA, and a second single-mode squeezer. Its action on an input coherent state with coherent amplitude $\alpha$ is written as~\cite{ralph2009}

\begin{equation}
|\alpha\rangle \longrightarrow S(r')T(g)S(r)|\alpha\rangle. \label{eq:ideal_device} 
\end{equation}
Here 
\begin{equation} 
S(r) = \exp\left[ \frac{r}{2} \left(a^2-a^{\dagger 2}\right) \right] \label{eq:squeezing_operator} 
\end{equation}
where \(r\) and \(r'\) are the squeezing parameters and $a$ and $a^{\dagger}$ are the creation and annihilation operators respectively.
The ideal NLA is described by 
\begin{equation}
T(g)=g^{\hat{n}}, \label{eq:ideal_nla} 
\end{equation} 
where \(\hat{n}\) is the photon-number operator and \(g\) is the gain of the NLA. Its action on a coherent state is
\begin{equation}
T(g)|\alpha\rangle \propto |g\alpha\rangle . 
\label{eq:ideal_nla_coherent} 
\end{equation}
The proportionality sign is required because \(T(g)\) is not a unitary operator. For \(g>1\), the transformation increases the coherent amplitude without adding the noise associated with a deterministic phase-insensitive amplifier.

The first squeezer changes both the displacement and the covariance matrix of the coherent state. After the ideal NLA, the state remains a minimum-uncertainty Gaussian state, but its two quadrature variances are unequal \cite{walk2013}. The second squeezing parameter can be chosen to restore the covariance matrix of a coherent state. This condition gives 

\begin{equation} r' = -\frac{1}{2} \ln \left[ \frac{ e^{-2r}+1-g^2(e^{-2r}-1) }{ e^{-2r}+1+g^2(e^{-2r}-1) } \right]. 
\label{eq:ideal_rp} 
\end{equation}
A derivation of the ideal transformation is given in
Appendix~\ref{app:ideal_derivation}.
With this choice of \(r'\), both quadratures are amplified by the same factor, \begin{equation} 
\boldsymbol{d}_{\mathrm{out}} = G\boldsymbol{d}_{\mathrm{in}}, \label{eq:ideal_effective_transformation} 
\end{equation} 
where 
\begin{equation} G = \frac{ 2g e^{-r} }{ \sqrt{ \left(e^{-2r}+1\right)^2 - g^4\left(e^{-2r}-1\right)^2 } }. \label{eq:ideal_gain} 
\end{equation}
The output has the covariance matrix of a coherent state and an
amplitude gain \(G\). Since the same gain multiplies both
quadratures in Eq.~(\ref{eq:ideal_effective_transformation}), the ideal
transformation applies to an arbitrary complex coherent amplitude
\(\alpha\), independent of its phase relative to the squeezing axes.

For \(r=0\), these expressions reduce to 

\begin{equation} 
r'=0, \qquad G=g, 
\end{equation}

and the device becomes a bare NLA. For nonzero squeezing, the hybrid device can give \(G>g\). It may therefore reach a given target gain using a lower NLA gain, suggesting a possible increase in the heralding probability. This result assumes an ideal NLA, which cannot operate over the full Fock space with a nonzero success probability. We therefore replace the ideal transformation with a finite-cutoff NLA and test whether the gain and probability advantages remain.

\subsection{Finite-cutoff noiseless linear amplification} \label{subsec:finite_cutoff}
We now replace the ideal NLA with a finite photon-number approximation. The squeezing operator obeys 
\begin{equation} S^{\dagger}(r)aS(r) = a\cosh r-a^{\dagger}\sinh r. \label{eq:squeezing_transformation} \end{equation}
In the calculations below, r and \(r'\) are taken to be real, thereby fixing the squeezing axes.
The coherent-state input is expanded as

\begin{equation}
|\alpha\rangle = e^{-|\alpha|^2/2} \sum_{k=0}^{\infty} \frac{\alpha^k}{\sqrt{k!}}|k\rangle .
\label{eq:coherent_fock_expansion} 
\end{equation} 
After the first squeezer, the state is written as 

\begin{equation}
S(r)|\alpha\rangle = \sum_{n=0}^{\infty}C_n|n\rangle, \label{eq:squeezed_input_expansion} 
\end{equation}

where 
\begin{equation} C_n = \langle n|S(r)|\alpha\rangle = e^{-|\alpha|^2/2} \sum_{k=0}^{\infty} \frac{\alpha^k}{\sqrt{k!}} \langle n|S(r)|k\rangle . \label{eq:Cn_general} 
\end{equation}

Thus, \(C_n\) contains both the coherent-state amplitude and the effect of the first squeezing operation.

Up to the common heralding factor, the cutoff-\(N\) NLA is
\begin{equation} T_N(g) = \sum_{n=0}^{N}g^n|n\rangle\langle n|. \label{eq:finite_nla} 
\end{equation} 
The unnormalized state after the NLA is 
\begin{equation} T_N(g)S(r)|\alpha\rangle = \sum_{n=0}^{N}g^nC_n|n\rangle . \label{eq:after_finite_nla} 
\end{equation}
The finite cutoff removes all components with \(n>N\). The
importance of these removed components depends on the input
amplitude and its phase relative to the fixed squeezing axes,
because the coefficients \(C_n\) are produced by the first
squeezer acting on the input coherent state. This truncation is the
source of the difference between the physical device and the ideal
hybrid transformation. Its phase dependence is examined in
Sec.~\ref{subsec:phase_dependence}.

The second squeezer gives the final unnormalized output 
\begin{equation} |\psi_{\mathrm{out}}^{(N)}\rangle = S(r')T_N(g)S(r)|\alpha\rangle = \sum_{m=0}^{\infty}C'_m|m\rangle , \label{eq:finite_output_state} 
\end{equation}
with 
\begin{equation} C'_m = \sum_{n=0}^{N} g^n C_n \langle m|S(r')|n\rangle . \label{eq:Cmprime_general}
\end{equation}

Here, \(N\) is the photon-number cutoff of the NLA, not of the final output state. The second squeezer transforms each retained \(|n\rangle\) into a superposition of Fock states with the same photon-number parity. It can produce components with \(m>N\). In the numerical calculations, the final output basis is chosen independently of N and sufficiently large to include the components generated by the second squeezer.

Equations~(\ref{eq:Cn_general}) and (\ref{eq:Cmprime_general}) define the calculation used for every NLA cutoff considered below. The coefficients are kept unnormalized at this stage. Their normalization and the definitions of gain, success probability, and fidelity are introduced in the following section.
The explicit low-order expansion of this matrix expression for
cutoffs 1--3 is given in Appendix~\ref{app:fock_expansion}.

\subsection{Performance measures} 
\label{subsec:performance}

The parameter \(g\) is the gain of the central finite-cutoff NLA,
whereas \(G\) denotes the effective gain of the complete hybrid
device. Because the finite cutoff distorts the output coefficients,
the gain inferred from different pairs of neighbouring Fock
coefficients need not be the same.

 We instead extract a lowest-order amplitude gain from the vacuum and one-photon coefficients.
The target coherent state has the expansion 
\begin{equation}
|G\alpha\rangle = e^{-|G\alpha|^2/2} \left( |0\rangle + G\alpha|1\rangle + \frac{(G\alpha)^2}{\sqrt{2}}|2\rangle + \cdots \right). \label{eq:target_coherent_expansion} 
\end{equation} 

Its neighboring Fock coefficients satisfy \(\frac{C'_{n+1}}{C'_n} = \frac{G\alpha}{\sqrt{n+1}}\). The gain inferred from each pair of neighboring coefficients is
\begin{equation}
G_{n,n+1}^{(N)} = \frac{\sqrt{n+1}}{\alpha} \frac{C'_{n+1}}{C'_n}. \label{eq:neighbouring_gain}
\end{equation}
For an exact amplified coherent state, all values of
\(G_{n,n+1}^{(N)}\) are equal to the same real gain \(G\). We use the
lowest-order value

\begin{equation} G_{01}^{(N)} = \frac{1}{\alpha} \frac{C'_1}{C'_0} \label{eq:G01_definition} 
\end{equation} 

as the effective gain of the finite-cutoff device. For the
phase-aligned inputs considered below, \(G_{01}^{(N)}\) is real, and
the target state is \(|G_{01}^{(N)}\alpha\rangle\). The higher-order
coefficient ratios are examined separately to determine whether this
lowest-order gain also describes the remaining output structure.

For the hybrid device, the total success probability of the cutoff-\(N\) NLA, including the \(N+1\) equivalent successful detection patterns, is \cite{winnel2020,guanzon2022} 
\begin{equation}
P_N^{\mathrm{hyb}} = \frac{N!} {(N+1)^{N-1}(1+g^2)^N} \sum_{n=0}^{N}g^{2n}|C_n|^2. 
\label{eq:hybrid_probability} 
\end{equation}
Here, \(g\) remains the gain of the central NLA, while the coefficients \(C_n\) include the first squeezing operation. The probability is independent of \(r'\), since the second squeezer is applied after the heralding measurement.

The corresponding bare NLA is chosen to have gain \(G\), so that it has the same lowest-order amplitude gain as the complete hybrid device. Its input coefficients are those of the original coherent state, giving
\begin{equation} P_N^{\mathrm{bare}} = \frac{N!} {(N+1)^{N-1}(1+G^2)^N} e^{-|\alpha|^2} \sum_{n=0}^{N} \frac{G^{2n}|\alpha|^{2n}}{n!}. \label{eq:bare_probability} 
\end{equation}

Equations~(\ref{eq:hybrid_probability}) and
(\ref{eq:bare_probability}) compare the hybrid and bare devices at
the same effective gain \(G\), rather than at the same internal NLA
gain.

The probability is calculated from the unnormalized state. The successful hybrid output is then normalized according to 
\begin{equation} |\psi_{\mathrm{out,norm}}^{(N)}\rangle = \frac{ \displaystyle\sum_m C'_m|m\rangle }{ \displaystyle\sqrt{\sum_m|C'_m|^2} }. \label{eq:normalized_output} 
\end{equation} 
Its fidelity with the target coherent state is

\begin{equation}
F_N^{\mathrm{hyb}} = \left| \langle G\alpha| \psi_{\mathrm{out,norm}}^{(N)} \rangle \right|^2. 
\label{eq:hybrid_fidelity}
\end{equation} 

The fidelity of the bare NLA is calculated against the same target state \(|G\alpha\rangle\).

\section{Finite-cutoff coherent-state amplification}
\label{sec:coherent_results}

\subsection{Emergence of squeezing-assisted gain}
\label{subsec:low_cutoff_gain}
\subsubsection{Cutoffs 1 and 2}
The low-cutoff sequence reveals how the ideal hybrid amplification mechanism emerges from the finite Fock space. We evaluate the coefficients using Eqs.~(\ref{eq:Cn_general}) and (\ref{eq:Cmprime_general}), and extract the effective gain from Eq.~(\ref{eq:G01_definition}).

\begin{figure*}[t]
    \centering
    \includegraphics[width=0.4\textwidth]
    {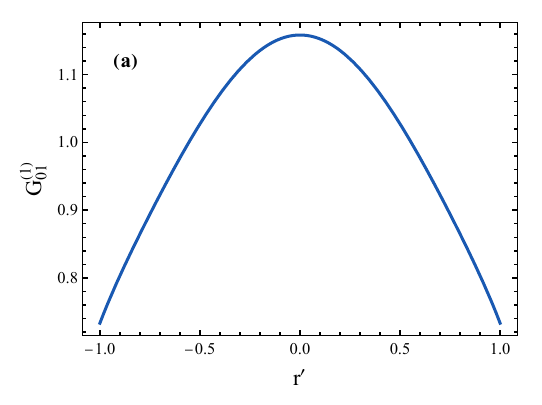}
    \hfill
    \includegraphics[width=0.4\textwidth]
    {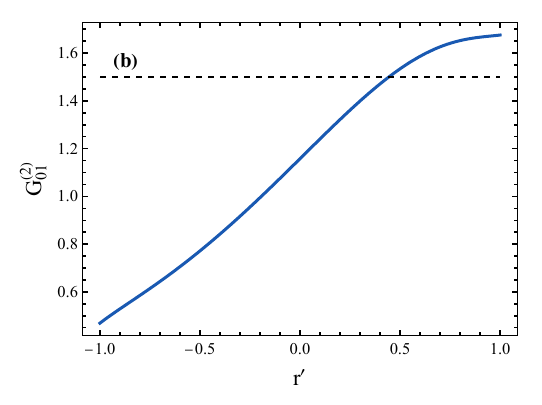}
    \caption{\label{fig:cutoff12_gain}
    Effective gain \(G_{01}^{(N)}\) as a function of the second
    squeezing parameter \(r'\) for (a) cutoff 1 and (b) cutoff 2.
    The parameters are \(r=0.75\), \(g=1.5\), and
    \(\alpha=0.1\). The dashed line denotes the internal NLA gain
    \(g=1.5\).}
\end{figure*}

At cutoff 1, only the vacuum and one-photon sectors pass through the NLA. No higher even- or odd-parity component is available to modify \(C'_0\) or \(C'_1\) after the second squeezer. As shown in Fig.~\ref{fig:cutoff12_gain}(a), \(G_{01}^{(1)}\) is maximal near \(r'=0\) and remains below the internal gain \(g=1.5\).

At cutoff 2, the NLA also retains the two-photon component. Under the second squeezer, this even-parity component contributes to \(C'_0\), but not to \(C'_1\). The resulting gain curve is shown in Fig.~\ref{fig:cutoff12_gain}(b). The additional contribution to \(C'_0\) changes the shape of the curve and allows \(G_{01}^{(2)}\) to exceed \(g\). However, the enhancement occurs for positive \(r'\), whereas the ideal compensation condition requires \(r'<0\). Cutoff 2 can increase \(G_{01}\), but it does not recover the gain mechanism of the ideal hybrid transformation.

\subsubsection{Cutoff 3}
Cutoff 3 introduces the first additional odd-parity component. Under
the second squeezer, the three-photon component contributes to
\(C'_1\), while the two-photon component contributes to \(C'_0\).
Consequently, both the numerator and denominator of
\(G_{01}^{(3)}\) are modified. This additional odd-parity channel is
the first qualitative change capable of reproducing the
squeezing-assisted gain mechanism of the ideal hybrid amplifier.

\begin{figure*}[t]
    \centering
    \includegraphics[width=0.4\textwidth]
    {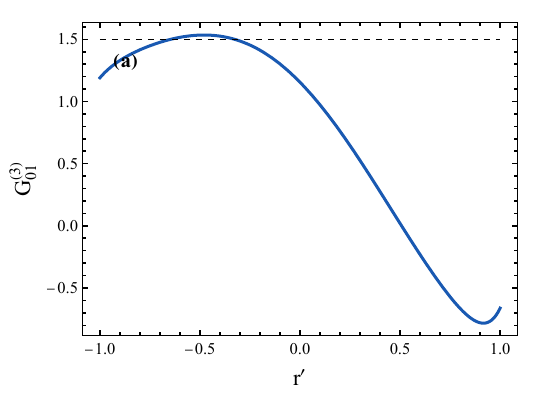}
    \hfill
    \includegraphics[width=0.4\textwidth]
    {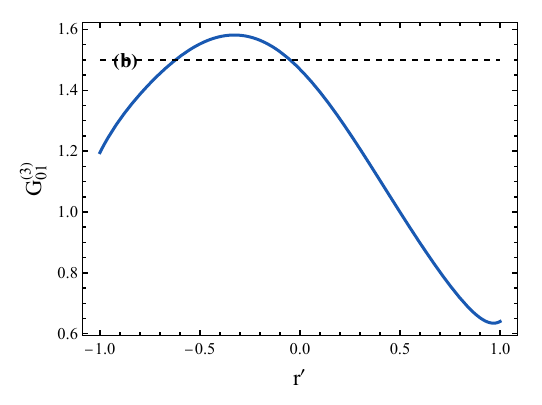}
    \caption{\label{fig:cutoff3_gain}
    Effective gain \(G_{01}^{(3)}\) as a function of \(r'\), with
    \(g=1.5\) and \(\alpha=0.1\), for (a) \(r=0.75\) and
    (b) \(r=0.20\). The dashed line denotes the internal NLA gain
    \(g=1.5\).}
\end{figure*}

Figure~\ref{fig:cutoff3_gain}(a) shows the cutoff-3 result for
\(r=0.75\), with \(g=1.5\) and \(\alpha=0.1\). The effective gain
slightly exceeds \(g\) for negative \(r'\), but only over a narrow
region, and the curve varies strongly with \(r'\). Reducing the first
squeezing parameter to \(r=0.20\), as shown in
Fig.~\ref{fig:cutoff3_gain}(b), produces a broader and more pronounced
gain enhancement in the negative-\(r'\) region. Cutoff 3 is therefore
the lowest cutoff for which squeezing-assisted amplification emerges
in the same squeezing direction as in the ideal transformation.

The value of \(r'\) that maximizes the finite-cutoff gain does not
generally coincide with the ideal compensation value. For
\(r=0.20\), the ideal transformation gives
\(r'\simeq-0.477\), whereas the cutoff-3 gain reaches its maximum near
\(r'\simeq-0.33\). This shift results from the removal of
photon-number components above the cutoff. Moreover, an enhanced
value of \(G_{01}^{(3)}\) constrains only the ratio of the vacuum and
one-photon coefficients and is not sufficient to establish
coherent-state amplification. The higher-order output coefficients
must therefore also reproduce the coherent-state structure, as
examined in the next subsection.

\subsection{Recovery of coherent-state structure}
\label{subsec:coherent_recovery}

The effective gain \(G_{01}^{(N)}\) constrains only the ratio of the one-photon and vacuum coefficients. Reproducing an amplified coherent state requires the remaining output coefficients to follow the same photon-number hierarchy. Expanding the target coherent state gives 
\begin{equation} |G\alpha\rangle = e^{-|G\alpha|^2/2} \sum_{m=0}^{\infty} \frac{(G\alpha)^m}{\sqrt{m!}}|m\rangle . \label{eq:target_coherent_full_expansion}
\end{equation}
The coefficients retained at cutoff \(N\) must satisfy \begin{equation} 
\frac{C'_m}{C'_1} = \frac{(G\alpha)^{m-1}}{\sqrt{m!}}, \qquad m=2,\ldots,N, \label{eq:coherent_hierarchy} 
\end{equation} 
where \(G=G_{01}^{(N)}\) is determined from the vacuum and one-photon coefficients. Equation~(\ref{eq:coherent_hierarchy}) tests whether the gain inferred from \(C'_1/C'_0\) is also consistent with the higher photon-number components. For each value of \(m\), we evaluate the two sides of Eq.~(\ref{eq:coherent_hierarchy}) as functions of \(r'\). Their intersection defines a coefficient-matching point \(r'_m\). A single condition being satisfied does not establish coherent-state amplification. Instead, the different values of \(r'_m\) must coincide, or form a sufficiently narrow cluster, for one second squeezer to reproduce the complete retained coefficient hierarchy. At cutoff 3, Eq.~(\ref{eq:coherent_hierarchy}) gives two independent conditions,
\begin{equation} 
\frac{C'_2}{C'_1} = \frac{G\alpha}{\sqrt{2}}, \qquad \frac{C'_3}{C'_1} = \frac{(G\alpha)^2}{\sqrt{6}}.
\label{eq:cutoff3_structure}
\end{equation} 
For \(r=0.20\), \(g=1.5\), and \(\alpha=0.1\), these conditions are satisfied at approximately \(r'_2=-0.38\) and \(r'_3=-0.34\), respectively. The two matching points are close but do not coincide. Cutoff 3 therefore provides only an approximate recovery of the coherent-state structure. Nevertheless, the proximity of the two solutions shows that the second squeezer can correct more than the lowest-order gain alone. Each increase in the cutoff introduces one additional condition in Eq.~(\ref{eq:coherent_hierarchy}). We evaluated these conditions for cutoffs 4--9. The matching points initially appear in neighbouring groups rather than converging immediately to a single value of \(r'\). In particular, conditions associated with adjacent photon-number orders tend to produce nearby intersections. These intermediate cutoffs thus reveal the progressive formation of a common compensation region, although the full set of retained coefficients cannot yet be associated with one sharply defined value of \(r'\). The convergence becomes substantially clearer at cutoff 11. We use \(\alpha=0.8\) so that the higher photon-number components make a non-negligible contribution, together with \(r=0.25\) and \(g=1.5\). The ten matching conditions corresponding to \(m=2,\ldots,11\) give 
\begin{equation}
\begin{split} \{r'_m\}_{m=2}^{11} =\{& -0.615,\,-0.623,\,-0.629,\,-0.619,\,-0.611,\\ & -0.620,\,-0.616,\,-0.630,\,-0.644,\,-0.635 \}. \label{eq:cutoff11_matching_points} 
\end{split}
\end{equation} All ten solutions lie within the narrow interval \begin{equation} 
-0.644 \lesssim r'_m \lesssim -0.611, 
\label{eq:cutoff11_matching_interval}
\end{equation} with a mean value of approximately \(-0.624\). We therefore select 
\begin{equation} 
r'=-0.62 \label{eq:cutoff11_working_point} 
\end{equation}
as the operating point for cutoff 11. At this point, the effective gain obtained from the vacuum and one-photon coefficients is
\begin{equation}
G_{01}^{(11)} \simeq 1.744, 
\label{eq:cutoff11_effective_gain} 
\end{equation} 
which exceeds the internal NLA gain \(g=1.5\). The clustering of all ten matching points shows that a single second squeezing parameter approximately restores the complete coefficient hierarchy retained at cutoff 11. The output is close to the target coherent state within the tested Fock subspace. The agreement is not exact, because the individual matching points remain slightly different and photon-number components above the cutoff are absent. The success probability and fidelity at this operating point are examined in the following subsection.
Throughout this analysis, \(\alpha\) is taken to be real and aligned
with one of the fixed squeezing axes. We first evaluate the
probability and fidelity at this operating point, and then examine
the effect of rotating the input phase.

\subsection{Success probability and fidelity}
\label{subsec:coherent_performance}

We now compare the physical performance of the hybrid device with that of a bare NLA. The comparison is made at the cutoff 11 operating point identified in the preceding subsection. For the hybrid device, we use
\begin{equation} r=0.25, \qquad g=1.5, \qquad r'=-0.62, \label{eq:cutoff11_hybrid_parameters} 
\end{equation} which gives an effective gain of \(G_{01}^{(11)}\simeq1.744\). The bare NLA is evaluated at the same cutoff and with its gain set to \(G=1.744\). The two devices therefore have the same target amplification, while the internal NLA gain required by the hybrid device is lower.

\begin{figure*}
\includegraphics[width=0.47\textwidth]
{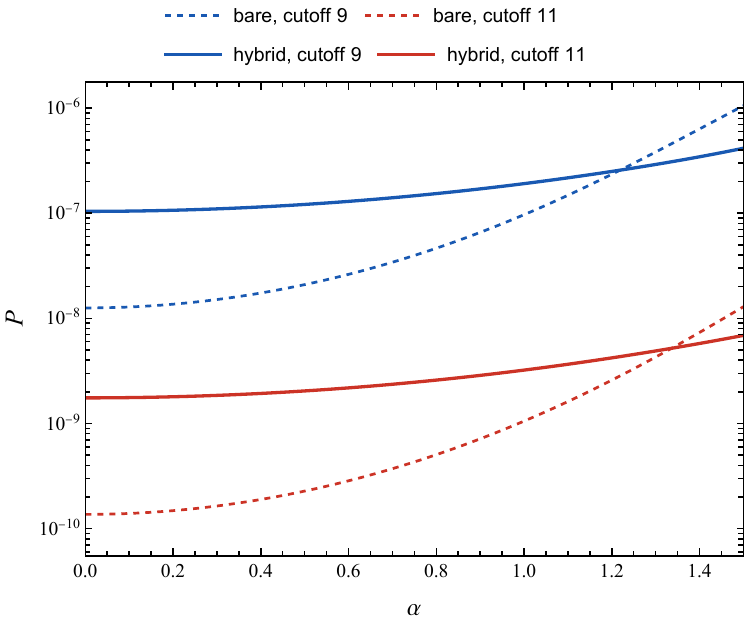}
\hfill
\includegraphics[width=0.47\textwidth]
{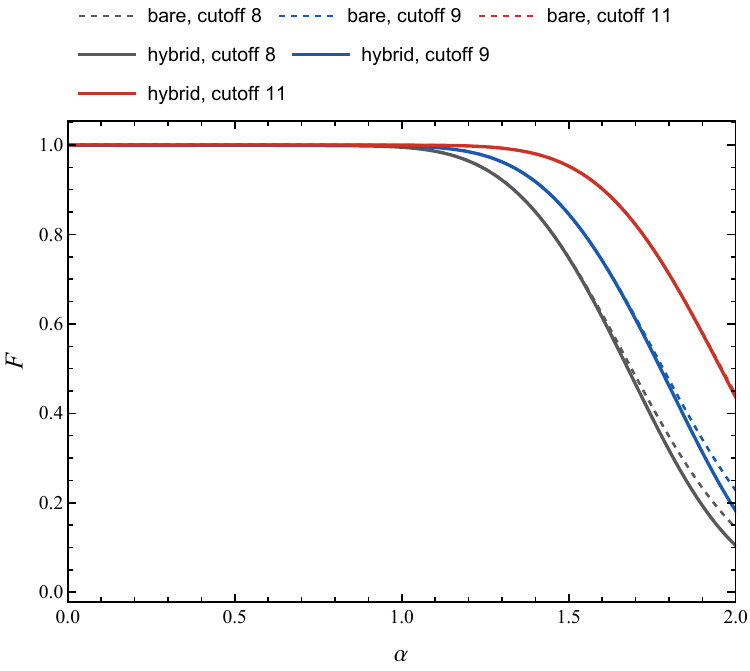}
\caption{\label{fig:coherent_performance}
Comparison of the bare NLA and the hybrid device. (a) Success
probability for cutoffs 9 and 11. Increasing the cutoff reduces the
absolute success probability, while the hybrid device gives a higher
probability than the bare NLA at each fixed cutoff before \(\alpha=1.4\). (b) Fidelity with
the target coherent state for cutoffs 8, 9, and 11. The hybrid
fidelity approaches that of the bare NLA as the cutoff is increased.
For each cutoff, the bare NLA is evaluated at the effective gain of
the corresponding hybrid device.}
\end{figure*}

Figure~\ref{fig:coherent_performance}(a) compares the success probabilities of the two devices as functions of the input coherent amplitude. The squeezing-assisted device has a higher heralding probability than the bare NLA at the matched effective gain. At the structure-matching amplitude \(\alpha=0.8\), we obtain approximately 
\begin{equation} 
P_{\mathrm{hyb}} \simeq 2.59\times10^{-9}, \qquad P_{\mathrm{bare}}
\simeq 5.07\times10^{-10}
\label{eq:cutoff11_probabilities}
\end{equation} 
The corresponding enhancement factor is 
\begin{equation} \frac{P_{\mathrm{hyb}}}{P_{\mathrm{bare}}} \simeq 5.1. \label{eq:cutoff11_probability_ratio}
\end{equation} 
Thus, the squeezing operations allow the target gain to be reached with a substantially larger success probability. The absolute probabilities nevertheless remain small, reflecting the experimental cost of implementing a high-cutoff NLA.

Figure~\ref{fig:coherent_performance}(b) shows the corresponding fidelities with the target coherent state \(|G\alpha\rangle\). At \(\alpha=0.8\), the fidelities are approximately \begin{equation}
F_{\mathrm{hyb}} \simeq 0.99978, \qquad F_{\mathrm{bare}} \simeq 0.999999. \label{eq:cutoff11_fidelities}
\end{equation}

The bare NLA gives the slightly higher fidelity, as expected from its direct truncated approximation to the target transformation. However, the fidelity reduction introduced by the squeezing-assisted scheme is of order \(10^{-4}\), while its success probability is increased by more than a factor of five.

These results establish the relevant tradeoff of the finite-cutoff hybrid amplifier. Squeezing does not improve the fidelity relative to a bare NLA at the same cutoff and target gain. Instead, it reduces the NLA gain required to obtain that target amplification and thereby increases the heralding probability, while introducing only a small reduction in output-state fidelity. At the operating point considered here, the device provides
a probability advantage without substantially degrading the
successful coherent-state output.

\begin{figure*}
\includegraphics[width=0.47\textwidth]
{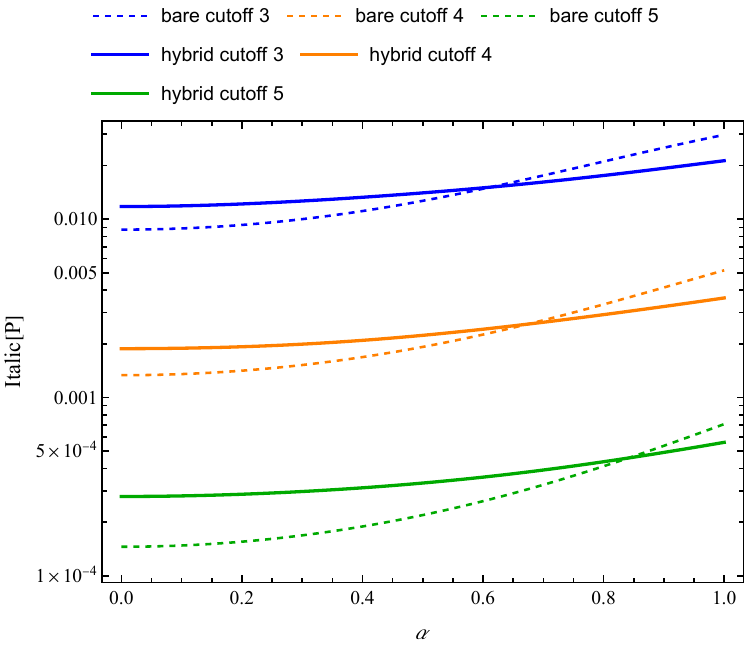}
\hfill
\includegraphics[width=0.47\textwidth]
{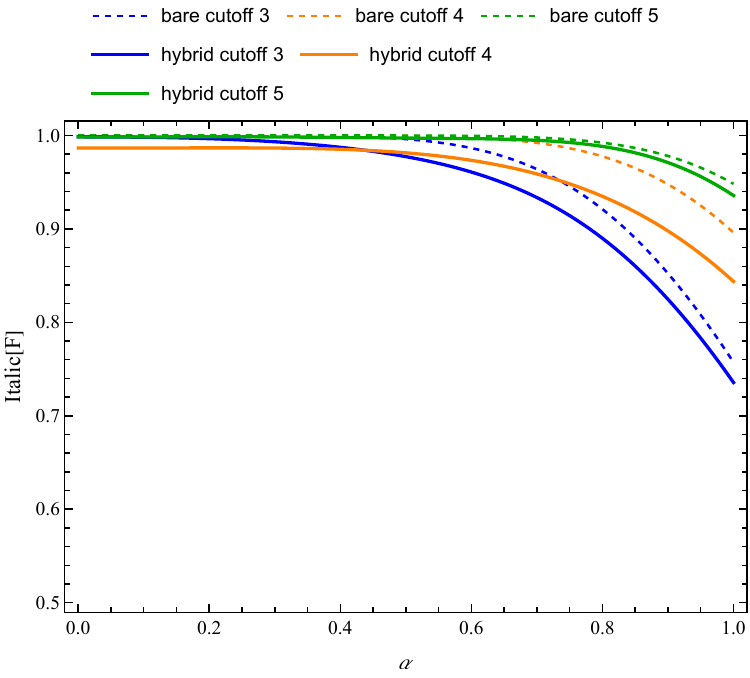}
\caption{\label{fig:lower_cutoff_performance}
Comparison of the hybrid amplifier and the corresponding bare NLA
at cutoffs 3, 4, and 5. For each cutoff, the second-squeezer
parameter \(r'\) is chosen at the maximum of the effective gain, and
the bare NLA is evaluated at that same gain. The common parameters
are \(r=0.2\) and \(g=1.5\). (a) Success probability and
(b) fidelity with the corresponding target coherent state as
functions of the real input amplitude \(\alpha\).}
\end{figure*}

To examine whether this advantage is already present at lower
cutoffs, Fig.~\ref{fig:lower_cutoff_performance} repeats the
comparison for cutoffs 3, 4, and 5. Each cutoff is evaluated at its
own maximum-gain operating point, and the corresponding bare NLA is
assigned the same effective gain. The comparison therefore shows how
the success-probability--fidelity tradeoff develops in the
lowest-order implementations of the hybrid scheme.

\subsection{Phase dependence of the finite-cutoff transformation} \label{subsec:phase_dependence}

The operating point above was obtained for the real input \(\alpha=0.8\). We now keep the same input magnitude and rotate the coherent amplitude to \(\alpha=0.8i\), while leaving the squeezing axes unchanged. For the ideal hybrid transformation, this rotation does not affect the amplification because both quadratures have the same gain. The cutoff 11 device, however, gives different results for the two input phases.

At \(r=0.25\), \(g=1.5\), and \(r'=-0.62\), the fidelities
with the target states \(|1.744\alpha\rangle\) are
\begin{equation}
F_{\mathrm{hyb}}(0.8)\simeq0.99978,
\qquad
F_{\mathrm{hyb}}(0.8i)\simeq0.823 .
\label{eq:real_imaginary_fidelity}
\end{equation}
At this point, the imaginary input gives
\(G_{01}^{(11)}\simeq1.723\), rather than the value \(1.744\)
obtained for the real input.
The difference originates before the second squeezer. After the first squeezer, the coefficients \(C_n(\alpha)\) depend on the phase of \(\alpha\) relative to the squeezing axes. The real and imaginary inputs have different photon-number distributions when they enter the NLA.

This difference can also be seen from the mean photon number after
the first squeezer. Using Eq.~(\ref{eq:squeezing_transformation}),
\begin{equation}
\begin{split}
    \langle \hat n\rangle_{\alpha=0.8}
    &=
    |\alpha|^2 e^{-2r}+\sinh^2r,\\
    \langle \hat n\rangle_{\alpha=0.8i}
    &=
    |\alpha|^2 e^{2r}+\sinh^2r .
\end{split}
\label{eq:real_imaginary_mean_photon}
\end{equation}
The imaginary input is displaced along the
anti-squeezed quadrature and has more weight at higher photon
numbers.

The effect of the cutoff can be obtained directly from the output
state in Eq.~(\ref{eq:finite_output_state}). Before normalization, the
photon-number components removed by the cutoff have total weight
\begin{equation}
    \sum_{n=N+1}^{\infty}g^{2n}|C_n(\alpha)|^2 .
    \label{eq:removed_components}
\end{equation}
Relative to the untruncated result, the removed fraction is
\begin{equation}
    \frac{
    \displaystyle\sum_{n=N+1}^{\infty}
    g^{2n}|C_n(\alpha)|^2
    }{
    \displaystyle\sum_{n=0}^{\infty}
    g^{2n}|C_n(\alpha)|^2
    } .
    \label{eq:removed_fraction}
\end{equation}

The retained and removed Fock components are orthogonal. Therefore,
the fidelity between the normalized cutoff-\(N\) output in
Eq.~(\ref{eq:finite_output_state}) and the corresponding untruncated output
is
\begin{equation}
    \frac{
    \left|
    \langle\psi_{\mathrm{out}}^{(\infty)}
    |\psi_{\mathrm{out}}^{(N)}\rangle
    \right|^2
    }{
    \langle\psi_{\mathrm{out}}^{(\infty)}
    |\psi_{\mathrm{out}}^{(\infty)}\rangle
    \langle\psi_{\mathrm{out}}^{(N)}
    |\psi_{\mathrm{out}}^{(N)}\rangle
    }
    =
    \frac{
    \displaystyle\sum_{n=0}^{N}g^{2n}|C_n(\alpha)|^2
    }{
    \displaystyle\sum_{n=0}^{\infty}g^{2n}|C_n(\alpha)|^2
    } .
    \label{eq:cutoff_output_overlap}
\end{equation}
Here, \(|\psi_{\mathrm{out}}^{(\infty)}\rangle\) denotes the
\(N\rightarrow\infty\) limit of Eq.~(\ref{eq:finite_output_state}). The
second squeezer does not alter this overlap because it is unitary.

For \(N=11\), \(r=0.25\), and \(g=1.5\), the fraction in
Eq.~(\ref{eq:removed_fraction}) is
\begin{equation}
\begin{split}
    \alpha=0.8 &: \qquad 2.19\times10^{-4},\\
    \alpha=0.8i &: \qquad 0.176 .
\end{split}
\label{eq:real_imaginary_removed_fraction}
\end{equation}
Equation~(\ref{eq:cutoff_output_overlap}) then gives overlap
fidelities of \(0.999781\) for \(\alpha=0.8\) and \(0.823951\)
for \(\alpha=0.8i\).

\begin{figure*}
\includegraphics[width=0.47\textwidth]
{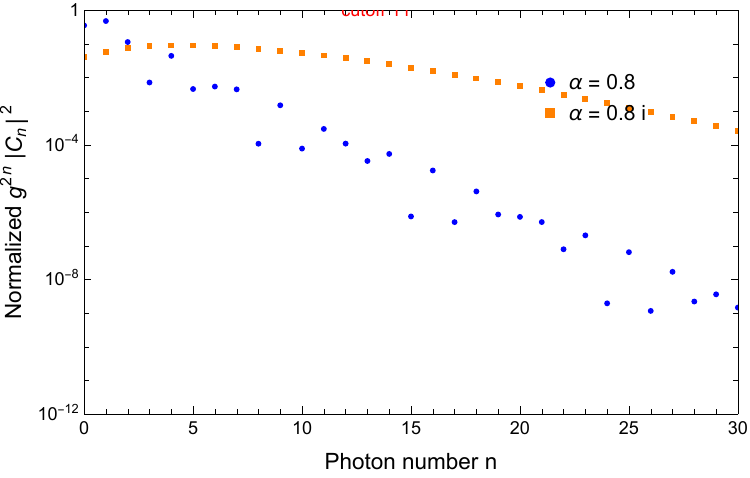}
\hfill
\includegraphics[width=0.47\textwidth]
{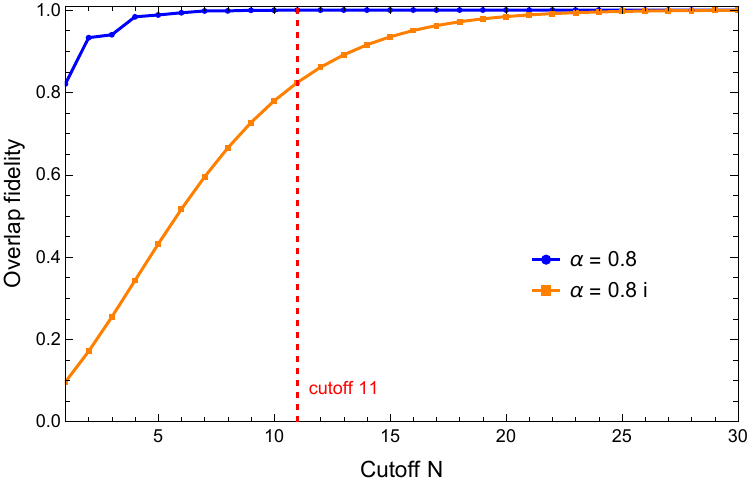}
\caption{\label{fig:phase_cutoff}
Effect of the finite photon-number cutoff for coherent inputs with
the same magnitude and orthogonal phases. The parameters are
\(r=0.25\) and \(g=1.5\). (a) Normalized photon-number weights
\(g^{2n}|C_n|^2\) after the first squeezer and NLA weighting.
The imaginary input has substantially more weight above cutoff 11.
(b) Fidelity between the cutoff-\(N\) output and the corresponding
untruncated output, obtained from
Eq.~(\ref{eq:cutoff_output_overlap}). The real input is already
converged at cutoff 11, whereas the imaginary input requires a
higher cutoff. The dashed lines indicate cutoff 11.}
\end{figure*}

Thus, cutoff 11 removes only about \(0.022\%\) of the
NLA-weighted norm for the real input, but about \(17.6\%\) for the
imaginary input. The factor \(g^{2n}\) makes the higher-photon-number components particularly important after the NLA.
The corresponding photon-number distributions and their convergence
with increasing cutoff are shown in Fig.~\ref{fig:phase_cutoff}.

This difference explains why the same value \(r'=-0.62\) works well
for the real input but not for the imaginary input. In the ideal
transformation, all photon-number components are retained and the
second squeezer restores the same gain in both quadratures. At finite
cutoff, the second squeezer can only act on the components that
remain after the NLA and cannot recover those removed above the
cutoff.

Changing \(r'\) can partially compensate for the altered coefficient
structure. Directly optimizing the imaginary-input fidelity gives
approximately
\begin{equation}
    r'=-0.442,\qquad
    G_{01}^{(11)}=1.711,\qquad
    F_{\mathrm{hyb}}=0.925 .
\label{eq:imaginary_optimized_point}
\end{equation}
This is an improvement over the result at \(r'=-0.62\), but it
remains below the real-input fidelity. Hence, the finite-cutoff device has a phase-dependent operating point and does not fully
recover the phase-independent behavior of the ideal hybrid
transformation at cutoff 11. However, in Fig.~\ref{fig:phase_cutoff} we plot fidelities for higher cutoffs and we see that eventually, approaching cutoff 30, the phase independent behavior of the ideal device is recovered.

\section{Application to a two-mode squeezed state}
\label{sec:tmsv_application}

\begin{figure*}
\centering

\resizebox{0.98\textwidth}{!}{
\begin{tikzpicture}[x=1cm,y=1cm]

\definecolor{lineblue}{RGB}{35,85,145}

\definecolor{squeezeFill}{RGB}{225,238,249}
\definecolor{squeezeEdge}{RGB}{55,112,164}

\definecolor{sourceFill}{RGB}{235,228,247}
\definecolor{sourceEdge}{RGB}{104,76,150}

\definecolor{nlaFill}{RGB}{251,229,224}
\definecolor{nlaEdge}{RGB}{184,72,54}

\definecolor{heraldGreen}{RGB}{43,126,89}

\tikzset{
    mode/.style={
        line width=1.15pt,
        draw=lineblue,
        -{Latex[length=2.4mm,width=1.7mm]}
    },
    device/.style={
        rounded corners=2pt,
        minimum height=12mm,
        align=center,
        font=\small,
        line width=0.85pt
    },
    squeeze/.style={
        device,
        draw=squeezeEdge,
        fill=squeezeFill,
        minimum width=16mm
    },
    nla/.style={
        device,
        draw=nlaEdge,
        fill=nlaFill,
        minimum width=24mm
    },
    source/.style={
        device,
        draw=sourceEdge,
        fill=sourceFill,
        minimum width=31mm,
        minimum height=25mm
    }
}

\node[
    anchor=east,
    font=\small
] (vacA) at (0,0.72)
{$|0\rangle_A$};

\node[
    anchor=east,
    font=\small
] (vacB) at (0,-0.72)
{$|0\rangle_B$};

\node[source] (s2) at (2.05,0)
{
    \textbf{TMSV source}\\[-1pt]
    $S_2(\chi)$
};

\draw[mode]
(vacA.east)
--
([yshift=7.2mm]s2.west);

\draw[mode]
(vacB.east)
--
([yshift=-7.2mm]s2.west);

\coordinate (refStart)
at ([yshift=7.2mm]s2.east);

\coordinate (sigStart)
at ([yshift=-7.2mm]s2.east);

\draw[mode]
(refStart)
--
(10.75,0.72);

\node[
    font=\scriptsize,
    fill=white,
    inner sep=1.5pt
]
at (4.05,0.72)
{mode $A$ (reference)};

\node[squeeze] (sr) at (4.65,-0.72)
{$S(r)$};

\node[nla] (nla) at (6.80,-0.72)
{
    \textbf{NLA}\\[-1pt]
    $\ T_N(g)$
};

\node[squeeze] (srp) at (9.05,-0.72)
{$S(r')$};

\draw[mode]
(sigStart)
--
(sr.west);


\draw[mode]
(sr.east)
--
(nla.west);

\draw[mode]
(nla.east)
--
(srp.west);

\draw[mode]
(srp.east)
--
(10.75,-0.72);

\node[
    font=\scriptsize,
    fill=white,
    inner sep=1.5pt
]
at (3.48,-0.47)
{mode $B$};

\draw[
    line width=0.9pt,
    draw=nlaEdge,
    -{Latex[length=2.1mm]}
]
(nla.south)
--
++(0,-0.48);

\node[
    below=1mm of nla,
    yshift=-5.5mm,
    rounded corners=2pt,
    draw=heraldGreen,
    fill=white,
    line width=0.7pt,
    inner xsep=4pt,
    inner ysep=2pt,
    font=\scriptsize,
    text=heraldGreen
]
{$\checkmark$ heralded success};

\draw[
    decorate,
    decoration={
        brace,
        amplitude=4.5pt
    },
    line width=0.8pt,
    draw=black
]
(11.05,-0.87)
--
(11.05,0.87);

\node[
    anchor=west,
    font=\small,
    align=left
]
at (11.30,0)
{
    $|\Psi_{\mathrm{out}}^{(N)}\rangle_{AB}$\\[-1pt]
    \scriptsize conditional output
};

\node[
    font=\scriptsize,
    align=center,
    text=black!78
]
at (2.05,-1.67)
{
    $|\Psi(\chi)\rangle_{AB}
    =
    \sqrt{1-\chi^2}
    \displaystyle\sum_{n=0}^{\infty}
    \chi^n|n\rangle_A|n\rangle_B$
};

\draw[
    densely dashed,
    rounded corners=3pt,
    draw=black!48,
    line width=0.7pt
]
(3.72,-1.44)
rectangle
(9.97,0.06);

\node[
    font=\scriptsize,
    fill=white,
    inner sep=1.5pt
]
at (6.85,-1.44)
{finite-cutoff hybrid amplifier};

\end{tikzpicture}
}

\caption{\label{fig:tmsv_setup}
Application of the finite-cutoff hybrid amplifier to one mode of a
two-mode squeezed vacuum state. Two vacuum modes enter the two-mode
squeezer \(S_2(\chi)\), producing the entangled state
\(\sqrt{1-\chi^2}\sum_{n=0}^{\infty}
\chi^n|n\rangle_A|n\rangle_B\).
Mode \(A\) is retained as the reference mode, while mode \(B\) is
transformed by the first single-mode squeezer \(S(r)\), the cutoff
\(N\) NLA \(T_N(g)\), and the second squeezer \(S(r')\).
A successful heralding outcome of the finite-cutoff NLA prepares the
conditional two-mode output
\( |\Psi_{\mathrm{out}}^{(N)}\rangle_{AB} \).}

\end{figure*}
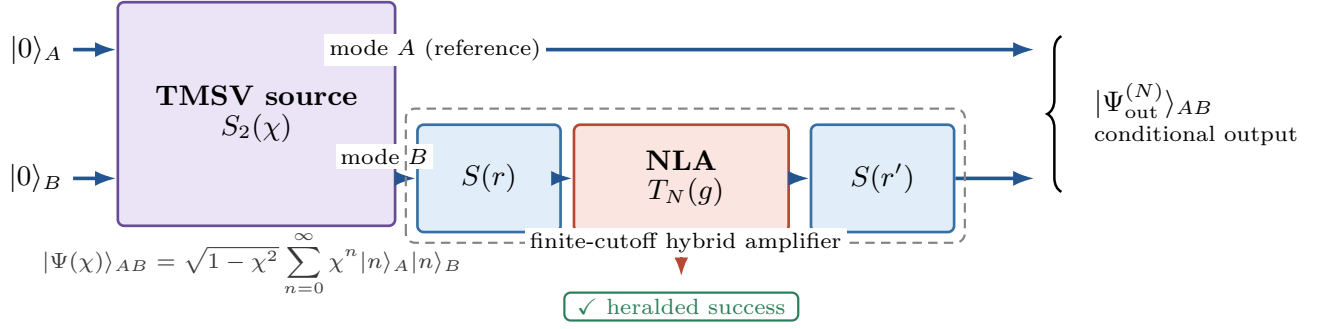

We now apply the finite-cutoff hybrid amplifier to one mode of a
two-mode squeezed vacuum state. The setup is shown in Fig.~\ref{fig:tmsv_setup}.

\subsection{Conditional output state}
\label{subsec:tmsv_output}
Two vacuum modes are first transformed by
a two-mode squeezer \(S_2(\chi)\), producing\cite{weedbrook2012}
\begin{equation}
    |\Psi(\chi)\rangle_{AB}
    =
    \sqrt{1-\chi^2}
    \sum_{n=0}^{\infty}
    \chi^n |n\rangle_A|n\rangle_B ,
    \label{eq:tmsv_input}
\end{equation}
where \(0\leq\chi<1\) determines the initial two-mode squeezing. In terms of the conventional two-mode squeezing parameter \(s\), one has \(\chi=\tanh s\).

Mode \(A\) is left unchanged, whereas mode \(B\) is transformed by the first single-mode squeezer, the cutoff \(N\) NLA, and the second single-mode squeezer. The unnormalized conditional output is
\begin{equation} 
|\widetilde{\Psi}_{\mathrm{out}}^{(N)}\rangle_{AB} = \left[ \mathbb{I}_A \otimes S(r')T_N(g)S(r) \right] |\Psi(\chi)\rangle_{AB}. \label{eq:tmsv_output_operator} 
\end{equation}
Expanding the transformation in the Fock basis gives 
\begin{equation} 
|\widetilde{\Psi}_{\mathrm{out}}^{(N)}\rangle_{AB} = \sqrt{1-\chi^2} \sum_{n=0}^{\infty} \sum_{m=0}^{\infty} \chi^n C_{m|n}^{\prime(N)} |n\rangle_A|m\rangle_B, 
\label{eq:tmsv_output_expansion} 
\end{equation}
where 
\begin{equation}
C_{m|n}^{\prime(N)} = \sum_{k=0}^{N} g^k \langle m|S(r')|k\rangle \langle k|S(r)|n\rangle . 
\label{eq:tmsv_output_coefficient} 
\end{equation}

The index \(n\) labels the photon number in the reference mode, whereas \(m\) labels the photon number in the processed output mode. Equation~(\ref{eq:tmsv_output_coefficient}) is the two-mode counterpart of the coherent-state coefficient transformation used in Sec.~\ref{sec:coherent_results}. The difference is that each photon-number component of mode \(B\) is now correlated with the corresponding component of mode \(A\). For comparison, an ideal NLA of effective gain \(G\), acting only on mode \(B\), gives \begin{equation} \left( \mathbb{I}_A\otimes G^{\hat n_B} \right) |\Psi(\chi)\rangle_{AB} \propto \sum_{n=0}^{\infty} (G\chi)^n |n\rangle_A|n\rangle_B. \label{eq:ideal_nla_tmsv} \end{equation} The ideal transformation therefore maps the TMSV parameter according to \begin{equation} \chi\longrightarrow G\chi, \label{eq:tmsv_parameter_amplification} \end{equation} subject to the normalizability condition \(G\chi<1\). This ideal result motivates the use of noiseless amplification in continuous-variable entanglement manipulation \cite{ralph2009,xiang2010,bernu2014}. A physical finite-cutoff device only approximates this transformation and operates conditionally, so its heralding probability must be evaluated explicitly.

The normalized conditional state is 
\begin{equation}
|\Psi_{\mathrm{out,norm}}^{(N)}\rangle_{AB} = \frac{ |\widetilde{\Psi}_{\mathrm{out}}^{(N)}\rangle_{AB} }{ \sqrt{ \langle \widetilde{\Psi}_{\mathrm{out}}^{(N)} | \widetilde{\Psi}_{\mathrm{out}}^{(N)} \rangle } }. \label{eq:tmsv_normalized_output} 
\end{equation}

\subsection{Heralding probability and output fidelity}
\label{subsec:tmsv_performance}

We evaluated the TMSV heralding probabilities for cutoffs from 3 to
11 and found a probability advantage over the corresponding bare NLA
throughout this range. We present cutoff 11 as the representative
case and examine whether its coherent-state operating point also
provides a useful conditional TMSV output.
 \begin{figure*}[t] 
 \centering \includegraphics[width=0.47\textwidth] {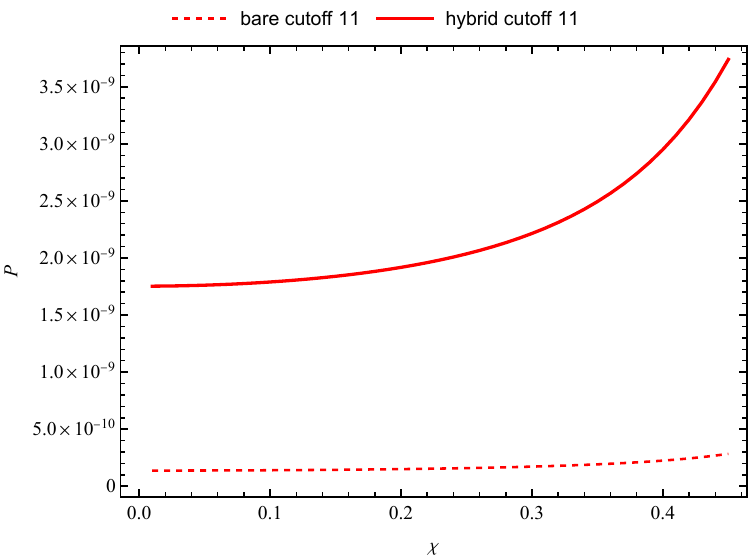} \hfill \includegraphics[width=0.47\textwidth] {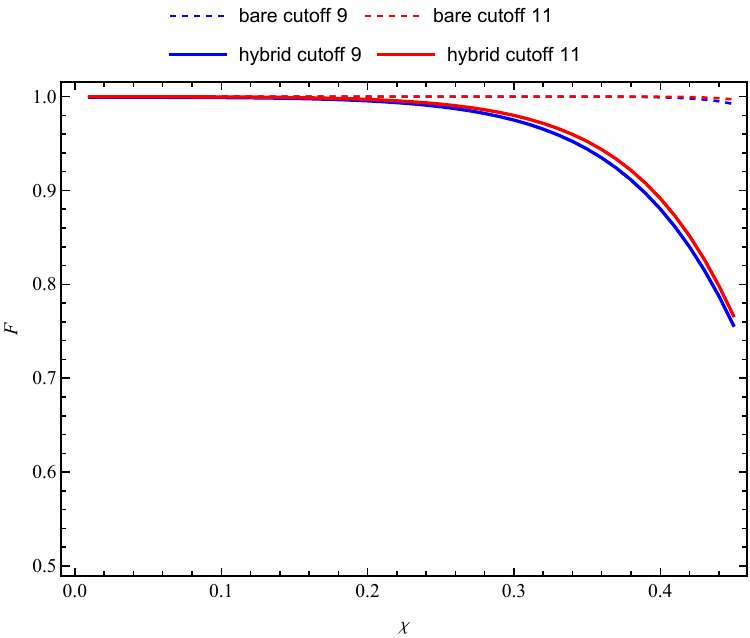} 
\caption{\label{fig:tmsv_performance}
Performance of the cutoff 11 hybrid device and bare NLA for a TMSV
input. The hybrid parameters are \(r=0.25\), \(g=1.5\), and
\(r'=-0.62\), giving the effective gain \(G=1.744\), while the bare
NLA operates directly at \(G=1.744\). (a) Heralding probability and
(b) fidelity with the output of an ideal, untruncated NLA as
functions of \(\chi\). The hybrid device has the higher success
probability, while its fidelity decreases more rapidly with
increasing \(\chi\).} 
\end{figure*}

We use cutoff 11 with \(r=0.25\), \(g=1.5\), and \(r'=-0.62\), for which the coherent-state analysis gives the effective gain \(G=1.744\). The corresponding bare NLA is evaluated with gain \(G=1.744\). For the TMSV input, the hybrid heralding probability is
\begin{equation} P_{\mathrm{h}}(\chi) = \frac{N!(1-\chi^2)} {(N+1)^{N-1}(1+g^2)^N} \sum_{n=0}^{\infty}\chi^{2n} \sum_{k=0}^{N} g^{2k} \left| \langle k|S(r)|n\rangle \right|^2 . 
\label{eq:tmsv_hybrid_probability} 
\end{equation} The sum over \(n\) accounts for the photon-number components of the reference mode, while the sum over \(k\) contains the components retained by the cutoff \(N\) NLA. The probability is independent of \(r'\).

For the bare NLA, \(S(r)=I\), and the probability reduces to 
\begin{equation} P_{N,\mathrm{bare}}(\chi) = \frac{N!(1-\chi^2)} {(N+1)^{N-1}(1+G^2)^N} \sum_{n=0}^{N}(G\chi)^{2n}. 
\label{eq:tmsv_bare_probability} 
\end{equation}

To assess the conditional output together with the success probability, we compare both devices with the normalized output of an ideal, untruncated NLA, \begin{equation} |\Psi_{\mathrm{ideal}}(G,\chi)\rangle_{AB} = \sqrt{1-G^2\chi^2} \sum_{n=0}^{\infty} (G\chi)^n|n\rangle_A|n\rangle_B , \label{eq:tmsv_ideal_normalized} 
\end{equation} 
where \(G\chi<1\). The normalized output of the bare cutoff-\(N\) NLA is \begin{equation}
|\Psi_{\mathrm{bare}}^{(N)}(\chi)\rangle_{AB} = \frac{ \displaystyle\sum_{n=0}^{N} (G\chi)^n|n\rangle_A|n\rangle_B }{ \displaystyle\sqrt{\sum_{n=0}^{N}(G\chi)^{2n}} }. \label{eq:tmsv_bare_normalized}
\end{equation} 

The hybrid and bare fidelities with the ideal output are

\begin{align}
F_{\mathrm{hyb}}^{(N)} &= \left| \left\langle \Psi_{\mathrm{ideal}} \middle| \Psi_{\mathrm{out,norm}}^{(N)} \right\rangle \right|^2, \label{eq:tmsv_hybrid_fidelity} \\ F_{\mathrm{bare}}^{(N)} &= \left| \left\langle \Psi_{\mathrm{ideal}} \middle| \Psi_{\mathrm{bare}}^{(N)} \right\rangle \right|^2. 
\label{eq:tmsv_bare_fidelity} 
\end{align} 
For the bare NLA, this gives
 \begin{equation} F_{\mathrm{bare}}^{(N)} = 1-(G\chi)^{2(N+1)}. \label{eq:tmsv_bare_fidelity_analytic}
 \end{equation}
Figure~\ref{fig:tmsv_performance} compares the probability and fidelity of the two devices. The hybrid device has the higher heralding probability because it produces the effective gain \(G=1.744\) using the lower internal NLA gain \(g=1.5\). The bare NLA more closely approximates the ideal output, while the hybrid fidelity decreases as \(\chi\) increases.

For \(G=1.744\), the ideal output exists only for
\(\chi<1/G\simeq0.573\). We restrict the calculation to
\(0\leq\chi\leq0.45\). The hybrid fidelity is approximately
\(0.980\) at \(\chi=0.3\), \(0.892\) at \(\chi=0.4\), and
\(0.767\) at \(\chi=0.45\). Thus, the hybrid device gives both high
fidelity and a probability advantage for weak TMSV inputs, but the
fidelity advantage is progressively lost as the input squeezing is
increased.

The origin of this reduction can be quantified from the
photon-number components presented to the finite-cutoff NLA. After
the first squeezer and the NLA weighting, the fractions of the norm
above cutoff 11 are approximately \(0.7\%\) at \(\chi=0.3\),
\(7.7\%\) at \(\chi=0.4\), and \(32.1\%\) at \(\chi=0.45\).
The increasing removal of these components accounts for the
pronounced fidelity reduction at larger \(\chi\).

The TMSV calculation result is not surprising given the phase dependence already identified at cutoff 11. In the fidelity plot of Figure~\ref{fig:tmsv_performance} we also plot the TMSV fidelity using a bare cutoff 9 device which shows similar behavior to the hybrid cutoff 11 device. We conclude that obtaining a gain advantage from the hybrid device whilst maintaining high fidelity would require higher cutoffs.

\section{Acknowledgement}
This work was
partially supported by the Australian Research Council Centre of Excellence for Quantum Computation and
Communication Technology (Project No.CE170100012).

\section{Conclusion}
\label{sec:conclusion}

We have studied how the amplification mechanism of an ideal
squeezer--NLA--squeezer device is modified when the central NLA is
replaced by a finite-cutoff quantum-scissors operation. The
low-cutoff sequence shows how the mechanism emerges from the
truncated Fock space. Cutoff 1 does not enhance the gain beyond that
of the central NLA, while cutoff 2 produces enhancement in the
opposite squeezing direction from the ideal transformation. Cutoff 3
is the first order at which the additional even- and odd-photon
components can produce gain enhancement in the required direction.

At cutoff 11, the matching conditions obtained from successive Fock
coefficients cluster around a common second-squeezing parameter. For
the phase-aligned input \(\alpha=0.8\), the operating point
\(r=0.25\), \(g=1.5\), and \(r'=-0.62\) gives an effective gain
\(G_{01}^{(11)}\simeq1.744\). The hybrid heralding probability is
approximately \(5.1\) times that of a bare cutoff 11 NLA operating at
the same effective gain, while the output fidelity remains
\(0.99978\). Squeezing can therefore convert a lower internal NLA
gain into a substantial probability advantage without significantly
degrading the successful output in this operating regime.

The finite-cutoff transformation does not, however, retain the
phase independence of the ideal hybrid amplifier. At the same
operating point, rotating the coherent amplitude from
\(\alpha=0.8\) to \(\alpha=0.8i\) reduces the fidelity to
approximately \(0.823\). After the first squeezer, the fractions of
the NLA-weighted norm above cutoff 11 are approximately \(0.022\%\)
and \(17.6\%\) for the real and imaginary inputs, respectively.
Reoptimizing the second squeezer improves the imaginary-input
fidelity to approximately \(0.925\), but does not recover the
phase-aligned result. The useful finite-cutoff operating point
therefore depends on the input phase relative to the squeezing axes.

We have also applied the device to one mode of a two-mode squeezed
vacuum state. The hybrid device retains its heralding advantage and
high fidelity for weak TMSV inputs. As the TMSV parameter increases,
a larger fraction of the NLA-weighted photon-number distribution lies
above the cutoff, and the hybrid fidelity decreases. 

Although our results indicate that impractically large cutoffs are required before the hybrid squeezing-NLA device behaves like the ideal device there may be restricted applications where phase dependent noiseless amplification is useful and an advantage may be seen. Beyond this, we believe the physics of how the ideal behavior emerges is interesting and may have applications, for example in quantum state engineering and the production of non-Gaussian quantum states.

\appendix

\section{Derivation of the ideal hybrid transformation}
\label{app:ideal_derivation}
This appendix gives the Gaussian-state derivation of the ideal results used in Sec.~\ref{sec:hybrid} \cite{mommers2017}. The displacement vector and covariance matrix of the input coherent state are 

\begin{equation} \boldsymbol{d}_{\mathrm{in}} = \begin{pmatrix} \langle x\rangle_{\mathrm{in}}\\ \langle p\rangle_{\mathrm{in}} \end{pmatrix}, \qquad \boldsymbol{V}_{\mathrm{in}} = \begin{pmatrix} 1&0\\ 0&1 \end{pmatrix}. \label{app:eq:input_moments} 
\end{equation} 
After the first squeezer, the displacement vector becomes 
\begin{equation} \boldsymbol{d}_{1} = \begin{pmatrix} e^{-r}\langle x\rangle_{\mathrm{in}}\\ e^{r}\langle p\rangle_{\mathrm{in}} \end{pmatrix}, \label{app:eq:first_displacement} 
\end{equation}

and the covariance matrix is 
\begin{equation} \boldsymbol{V}_{1} = \begin{pmatrix} e^{-2r}&0\\ 0&e^{2r} \end{pmatrix}. \label{app:eq:first_covariance} 
\end{equation}

For a Gaussian state with covariance matrix \(\operatorname{diag}(V_x,V_p)\), the ideal NLA transforms the displacement vector as \cite{walk2013} 

\begin{equation} \boldsymbol{d} \longrightarrow \begin{pmatrix} \displaystyle \frac{2g\langle x\rangle} {V_x+1-g^2(V_x-1)} \\[8pt] \displaystyle \frac{2g\langle p\rangle} {V_p+1-g^2(V_p-1)} \end{pmatrix}, \label{app:eq:nla_displacement} 
\end{equation}

while its covariance matrix becomes

\begin{equation} \boldsymbol{V} \longrightarrow \begin{pmatrix} \displaystyle \frac{V_x+1+g^2(V_x-1)} {V_x+1-g^2(V_x-1)} &0 \\[10pt] 0& \displaystyle \frac{V_p+1+g^2(V_p-1)} {V_p+1-g^2(V_p-1)} \end{pmatrix}. \label{app:eq:nla_covariance} 
\end{equation}

Using \(V_x=e^{-2r}\) and \(V_p=e^{2r}\), the displacement vector after the NLA is 

\begin{equation} \boldsymbol{d}_{2} = \begin{pmatrix} \displaystyle \frac{2g e^{-r}\langle x\rangle_{\mathrm{in}}} {e^{-2r}+1-g^2(e^{-2r}-1)} \\[10pt] \displaystyle \frac{2g e^{r}\langle p\rangle_{\mathrm{in}}} {e^{2r}+1-g^2(e^{2r}-1)} \end{pmatrix}. \label{app:eq:after_nla_displacement} \end{equation} The corresponding covariance matrix is 
\begin{equation} 
\boldsymbol{V}_{2} = \begin{pmatrix} \displaystyle \frac{e^{-2r}+1+g^2(e^{-2r}-1)} {e^{-2r}+1-g^2(e^{-2r}-1)} &0 \\[12pt] 0& \displaystyle \frac{e^{2r}+1+g^2(e^{2r}-1)} {e^{2r}+1-g^2(e^{2r}-1)} \end{pmatrix}. \label{app:eq:after_nla_covariance} 
\end{equation} 

The two diagonal elements of Eq.~(\ref{app:eq:after_nla_covariance}) are reciprocal. The state therefore remains a minimum-uncertainty Gaussian state, although its quadrature variances are unequal. The second squeezer transforms the covariance matrix according to 

\begin{equation}
\boldsymbol{V}_{\mathrm{out}} = \begin{pmatrix} e^{-2r'}&0\\ 0&e^{2r'} \end{pmatrix} \boldsymbol{V}_{2}. \label{app:eq:second_covariance} \end{equation}

Requiring the output covariance matrix to satisfy 
\begin{equation}
\boldsymbol{V}_{\mathrm{out}} = \begin{pmatrix} 1&0\\ 0&1 \end{pmatrix} \label{app:eq:output_covariance} 
\end{equation}

gives 
\begin{equation}
r' = -\frac{1}{2} \ln \left[ \frac{ e^{-2r}+1-g^2(e^{-2r}-1) }{ e^{-2r}+1+g^2(e^{-2r}-1) } \right], \label{app:eq:rp_result}
\end{equation}

which is Eq.~(\ref{eq:ideal_rp}) in the main text. The output displacement vector is 
\begin{equation}
\boldsymbol{d}_{\mathrm{out}} = \begin{pmatrix} e^{-r'}&0\\ 0&e^{r'} \end{pmatrix} \boldsymbol{d}_{2}. \label{app:eq:output_displacement} \end{equation}

Substituting Eq.~(\ref{app:eq:rp_result}) into Eq.~(\ref{app:eq:output_displacement}) gives the same amplification factor for both quadratures: 
\begin{equation} \boldsymbol{d}_{\mathrm{out}} = G\boldsymbol{d}_{\mathrm{in}}, \label{app:eq:effective_transformation}
\end{equation}
where 
\begin{equation} G = \frac{ 2g e^{-r} }{ \sqrt{ \left(e^{-2r}+1\right)^2 - g^4\left(e^{-2r}-1\right)^2 } }. \label{app:eq:gain_result} 
\end{equation}
This reproduces Eq.~(\ref{eq:ideal_gain}) in the main text. For \(r=0\), the result reduces to \(r'=0\) and \(G=g\), as expected for a bare NLA.

\section{Fock-basis expansion of the finite-cutoff transformation}
\label{app:fock_expansion}

This appendix derives the squeezing-matrix elements used in the
analytic cutoff 1--3 calculation. The general finite-cutoff output
is already given in the main text, so only the matrix elements
required to evaluate that expression are derived here.

The squeezing operator in Eq.~(\ref{eq:squeezing_operator}) can also be written in the
disentangled form
\begin{equation}
\begin{aligned}
 S(x)
 &=
 \exp\left[
 -\frac{\tanh x}{2}a^{\dagger 2}
 \right]
 (\operatorname{sech}x)^{a^\dagger a+1/2}\\
 &\quad\times
 \exp\left[
 \frac{\tanh x}{2}a^2
 \right].
\end{aligned}
\label{eq:app_disentangled_squeezer}
\end{equation}
This form is convenient because the annihilation exponential acts
first on the input Fock state, the middle factor is diagonal in the
Fock basis, and the creation exponential then generates states of
the same photon-number parity. Consequently,
\begin{equation}
 \langle m|S(x)|n\rangle=0
 \qquad\text{when \(m\) and \(n\) have opposite parity.}
 \label{eq:app_parity_rule}
\end{equation}

For the vacuum matrix element, the rightmost exponential acts
trivially on \(|0\rangle\). The vacuum component of the leftmost
exponential is its zeroth-order term, giving
\begin{equation}
 \langle0|S(x)|0\rangle
 =
 \sqrt{\operatorname{sech}x}.
 \label{eq:app_s00_exact}
\end{equation}
Expanding this exact expression about \(x=0\) gives
\begin{equation}
\begin{aligned}
 \langle0|S(x)|0\rangle
 &=
 1-\frac{x^2}{4}
 +\frac{7x^4}{96}
 -\frac{139x^6}{5760}\\
 &\quad
 +\frac{5473x^8}{645120}
 -\frac{51103x^{10}}{16588800}
 +O(x^{12}).
\end{aligned}
\label{eq:app_s00_series}
\end{equation}

For the one-photon diagonal element, the annihilation and creation
exponentials cannot contribute without changing the photon number.
The diagonal factor therefore gives
\begin{equation}
 \langle1|S(x)|1\rangle
 =
 (\operatorname{sech}x)^{3/2}.
 \label{eq:app_s11_exact}
\end{equation}
Its expansion is
\begin{equation}
\begin{aligned}
 \langle1|S(x)|1\rangle
 &=
 1-\frac{3x^2}{4}
 +\frac{13x^4}{32}
 -\frac{379x^6}{1920}\\
 &\quad
 +\frac{19627x^8}{215040}
 -\frac{1588441x^{10}}{38707200}
 +O(x^{12}).
\end{aligned}
\label{eq:app_s11_series}
\end{equation}

The two-photon component generated from the vacuum comes from the
term linear in \(a^{\dagger 2}\) in the first exponential of
Eq.~\eqref{eq:app_disentangled_squeezer}. Using
\(a^{\dagger 2}|0\rangle=\sqrt{2}|2\rangle\), we obtain
\begin{equation}
 \langle2|S(x)|0\rangle
 =
 -\frac{\tanh x}{\sqrt{2}}
 \sqrt{\operatorname{sech}x}.
 \label{eq:app_s20_exact}
\end{equation}
Therefore,
\begin{equation}
\begin{aligned}
 \langle2|S(x)|0\rangle
 &=
 -\frac{x}{\sqrt{2}}
 +\frac{7x^3}{12\sqrt{2}}
 -\frac{139x^5}{480\sqrt{2}}\\
 &\quad
 +\frac{5473x^7}{40320\sqrt{2}}
 -\frac{51103x^9}{829440\sqrt{2}}
 +O(x^{11}).
\end{aligned}
\label{eq:app_s20_series}
\end{equation}

Because \(S^\dagger(x)=S(-x)\), the reversed matrix element has the
opposite sign:
\begin{equation}
 \langle0|S(x)|2\rangle
 =
 \frac{\tanh x}{\sqrt{2}}
 \sqrt{\operatorname{sech}x}.
 \label{eq:app_s02_exact}
\end{equation}
Its expansion is
\begin{equation}
\begin{aligned}
 \langle0|S(x)|2\rangle
 &=
 \frac{x}{\sqrt{2}}
 -\frac{7x^3}{12\sqrt{2}}
 +\frac{139x^5}{480\sqrt{2}}\\
 &\quad
 -\frac{5473x^7}{40320\sqrt{2}}
 +\frac{51103x^9}{829440\sqrt{2}}
 +O(x^{11}).
\end{aligned}
\label{eq:app_s02_series}
\end{equation}

Similarly, the three-photon component generated from
\(|1\rangle\) follows from
\(a^{\dagger 2}|1\rangle=\sqrt{6}|3\rangle\). This gives
\begin{equation}
 \langle3|S(x)|1\rangle
 =
 -\frac{\sqrt{6}}{2}
 \tanh x\,(\operatorname{sech}x)^{3/2}.
 \label{eq:app_s31_exact}
\end{equation}
Expanding about \(x=0\),
\begin{equation}
\begin{aligned}
 \langle3|S(x)|1\rangle
 &=
 -\frac{\sqrt{6}}{2}x
 +\frac{13\sqrt{6}}{24}x^3
 -\frac{379\sqrt{6}}{960}x^5\\
 &\quad
 +\frac{19627\sqrt{6}}{80640}x^7\\
 &\quad
 -\frac{1588441\sqrt{6}}{11612160}x^9 +O(x^{11}).
\end{aligned}
\label{eq:app_s31_series}
\end{equation}

The reversed matrix element is
\begin{equation}
 \langle1|S(x)|3\rangle
 =
 \frac{\sqrt{6}}{2}
 \tanh x\,(\operatorname{sech}x)^{3/2},
 \label{eq:app_s13_exact}
\end{equation}
with the expansion
\begin{multline}
\langle1|S(x)|3\rangle
=
\frac{\sqrt{6}}{2}x
-\frac{13\sqrt{6}}{24}x^3
+\frac{379\sqrt{6}}{960}x^5
\\
-\frac{19627\sqrt{6}}{80640}x^7
+\frac{1588441\sqrt{6}}{11612160}x^9
\\
-\frac{185176333\sqrt{6}}{2554675200}x^{11}
+O(x^{13}).
\label{eq:app_s13_series}
\end{multline}

Substituting these matrix elements into the cutoff 1--3 expressions
in the main text gives the corresponding analytic gain curves. The
series expansions are used only to display the low-cutoff mechanism
explicitly. The cutoff 4--11 results are calculated using the full
numerical matrix exponential of the squeezing operator. In those
calculations, the NLA input is truncated at the physical cutoff
\(N\), whereas the output of the second squeezer is retained in the
complete numerical basis.

\bibliography{apssamp}

\end{document}